\documentclass[aps,prd,showpacs,nofootinbib,superscriptaddress,preprint,tightenlines]{revtex4}

\usepackage{color}

\usepackage{graphicx}
\usepackage[centertags]{amsmath}
\usepackage{amsfonts}
\usepackage{amssymb}
\usepackage{amsthm}
\usepackage{newlfont}

\def\bea{\begin{eqnarray}}
\def\eea{\end{eqnarray}}
\def\be{\begin{equation}}
\def\ee{\end{equation}}

\usepackage{graphicx}

\def\be{\nopagebreak[3]\begin{equation}}
\def\ee{\end{equation}}
\def\ba{\nopagebreak[3]\begin{eqnarray}}
\def\ea{\end{eqnarray}}

\def\R{\mathbb{R}}
\def\C{\mathbb{C}}

\newcommand{\teta}{\rlap{\lower2ex\hbox{$\,\tilde{}$}}\eta{}}

\def\R{\mathbb{R}}

\newcommand{\f}{\frac}

\usepackage{enumerate}
\DeclareMathOperator*{\sgn}{sgn}

\usepackage{colordvi}

\usepackage[centertags]{amsmath}
\usepackage{amsfonts}
\usepackage{amssymb}
\usepackage{amsthm}
\usepackage{dsfont}
\usepackage{newlfont}
\usepackage{hyperref}
\def\be{\begin{equation}}
\def\ee{\end{equation}}
\def\dif{\textrm{d}}

\def\f{\frac}

\begin{document}
\preprint{\vbox{\baselineskip=12pt \rightline{IGC-13/7-5}
}}

\title{Loop quantum cosmology of $k=1$ FLRW:\\ Effect of inverse volume corrections}
\author{Alejandro Corichi}\email{corichi@matmor.unam.mx}
\affiliation{Centro de Ciencias Matem\'aticas, Universidad Nacional Aut\'onoma de
M\'exico, UNAM-Campus Morelia, A. Postal 61-3, Morelia, Michoac\'an 58090,
Mexico}
\affiliation{Center for Fundamental Theory, Institute for Gravitation and the Cosmos,
Pennsylvania State University, University Park
PA 16802, USA}
\author{Asieh Karami}
\email{karami@matmor.unam.mx}
\affiliation{Instituto de F\'{\i}sica y
Matem\'aticas,  Universidad Michoacana de San Nicol\'as de
Hidalgo, Morelia, Michoac\'an, Mexico}
\affiliation{Centro de Ciencias Matem\'aticas, Universidad Nacional Aut\'onoma de
M\'exico, UNAM-Campus Morelia, A. Postal 61-3, Morelia, Michoac\'an 58090,
Mexico}

\begin{abstract}
We consider the $k$=1 Friedman-Lemaitre-Robertson-Walker (FLRW) model within loop quantum cosmology (LQC), from the perspective of the two available quantization prescriptions. We focus our attention on the existence of
the so called  `inverse corrections' in the quantization process. We derive the corresponding
quantum constraint operators, and study in detail the issue of the self-adjointness of the
quantum Hamiltonian constraint for two different quantizations based on open holonomies. Furthermore, we analize in detail the resulting effective theories, paying special attention to issues such as the boundedness of different scalar observables such as energy density and expansion. We show that the inclusion of the inverse corrections solves one of the main undesirable features of the previous formalism, namely the unboundedness of the energy density and expansion near the bounce on the full (effective) phase space. We briefly comment on possible consequences of this new feature.
 \end{abstract}

\maketitle

\section{Introduction}
\label{sec:0}

One of the main challenges that any quantum gravity candidate theory faces is to address the issue of gravitational singularities. In the context of cosmological models, this means facing the big bang/crunch singularities head on. Experience with other instances in physics suggests that a consistent quantum theory should resolve the singularity and replace it by something else. Loop quantum cosmology (LQC) \cite{lqc, AA, aa:ps, ia:ac}, a symmetry reduced model closely related to loop quantum gravity \cite{lqg}, has provided very clear answers for certain models. The most widely studied model, namely a $k$=0 FLRW cosmology, has fulfilled that expectation with flying colors; it has been shown that the big bang/crunch singularity is replaced by a big bounce connecting a contracting and an expanding branch \cite{aps2,slqc}. This bounce was been shown to be generic and to provide an absolute bound on the matter density for all physical quantum states. Semiclassical states have also been studied in detail and it has been shown that semiclassicality is preserved across the bounce \cite{cs:prl,recall,CM}. A very interesting feature of the quantum dynamics of semiclassical states is that their dynamics can be very well approximated by a set of `effective equations' that capture the main contribution of the    `quantum loop effects'. These equations are very well approximated by the Einstein's equations at small densities/curvature, but differ considerably from them near the Planck scale \cite{aps2,singh:vander}. Consistency of the effective description has allowed to discern, from a wide possibility of quantum theories, only one that was introduced by different geometric considerations \cite{cs:unique,aps2}.   One can also show that the effective equations describe with very good accuracy the main
features of the quantum dynamics for appropriately defined physical semiclassical states \cite{CM}.

For other isotropic models, such as the closed $k$=1 FRLW model here considered, we do not have
that much wealth of information. In \cite{apsv}, the authors considered the loop quantization
of the model, and performed intensive numerical evolutions of semiclassical states that were shown to follow the dynamics dictated by the effective equations (studied in \cite{pol}). In \cite{closed} the quantum Hamiltonian was proven to be essentially self-adjoint. A new quantization, mimicking the strategy employed in (spatially curved) anisotropic models in LQC \cite{bianchiII, bianchiIX} was put forward in \cite{ck2}.
The idea is to define the quantum operator approximating the connection by means of {\em open holonomies}. To emphasize this feature, this quantization method was dubbed the {\em connection based quantization} and the original one introduced in \cite{apsv} was named the {\em curvature based quantization}.
In \cite{ck2} it was shown that the effective theory of the new quantization provides a very different cyclic universe scenario. Instead of having a succession of bounces and recollapses as in \cite{apsv}, one now has two different kind of bounces, and recollapses. The universe is still cyclic but only after going through two phases of bounce and recollapse.  However, some issues were left unresolved. For instance. in most of the analysis, the so called {\em inverse volume corrections} of LQC were largely ignored, as was the self adjointness of the quantum operator. Furthermore, it is known that some attractive features of the effective dynamics
of the flat FLRW ere absent in the dynamics of the new theory. In particular, neither the expansion nor the energy density are absolutely bounded, while in the curvature based quantization it is only the expansion that is bounded \cite{ck2,gupt:singh}.

The purpose of this paper is to extend the results of \cite{ck2} and fill all these gaps. First, we consider from the very beginning the inverse volume corrections, for both quantization methods. Second, we analyse in detail the self-adjointness of the resulting quantum operators for the connection based quantization. We show that the operators are indeed essentially self-adjoint so one can have confidence that the evolution of the states (yet to be done) will be
well defined. Finally, we analyse the corresponding effective theories, for both quantization methods, with the modifications coming from the inverse volume corrections. Interestingly, the undesirable features disappear; for both quantization methods, the expansion and the
energy density are absolutely bounded in the whole (effective) phase space. This feature might have important ramifications when computing the probability of inflation, as was done in \cite{inflation0,inflation1} for the flat case.

The structure of the paper is as follows. In Sec.~\ref{sec:1} we introduce some preliminaries regarding the $k$=1 FLRW model, following \cite{ck2}. In Sec.~\ref{sec:2} we introduce the quantization for both cases, including the inverse corrections. In the case of the connection based quantization we introduce two different factor orderings. The issue of self-adjointness is considered in Sec.~\ref{sec:3}. The effective theories for both quantizations is studied in Sec.~\ref{sec:4}. We end with some discussion in Sec.~\ref{sec:5}.

\section{Preliminaries: The closed FLRW Model}
\label{sec:1}

In this section we shall recall some preliminary material regarding the spatially closed model. We shall follow the notation of \cite{ck2}.
In this model, as in all FLRW models, the spacetime is taken of the form $M=\Sigma\times\mathbb{R}$, where the underlying spatial {\em homogeneous} 3-manifold $\Sigma$ is topologically a 3-sphere $\mathbb S^3$ which can be identified with the symmetry group SU(2). One can naturally define fiducial co-frames ${}^o\omega_a^i$ on $\Sigma$ via,
\be
g^{-1}\dif g:={}^o\omega={}^o\omega^i\tau_i
\ee
where $g$ is an element of SU(2). We denote by ${}^oe_i^a$ the corresponding dual frames
such that ${}^oe_i^a{}^o\omega^j_a=\delta_i^j$ and ${}^oe_i^a{}^o\omega^i_b=\delta_b^a$.
From the above definition for natural left-invariant 1-form ${}^o\omega,$ it follows that the
frames ${}^oe_i^a$ and the co-frames ${}^o\omega^i_a$ satisfy the Maurer-Cartan equations
\be
\label{MCE}
\dif {}^o\omega^i+\frac{1}{2}{}^o\epsilon^i_{\ jk}{}^o\omega^j\wedge{}^o\omega^k=0\ \ \ \textrm{and}\ \ \ [{}^oe_i,{}^oe_j]={}^o\epsilon_{ij}^{\ \ k}{}^oe_k
\ee
Where ${}^o\epsilon_{ijk}$ is the completely antisymmetric tensor defined such that ${}^o\epsilon_{123}=1$. We will use ${}^o\omega^i$ and ${}^oe_i$ as our fiducial frames and co-frames.

The fiducial metric on $\Sigma$ is ${}^oq_{ab}={}^o\omega_a^i{}^o\omega_b^jk_{ij}$ where $k_{ij}$ is the Killing-Cartan metric on su(2). The fiducial metric ${}^oq_{ab}$ is the metric on the `round' 3-sphere with radius $a_o=2$.
The volume of $\Sigma$, with respect to the metric ${}^oq_{ab}$ is $V_o=2\pi^2a_o^3$. We define the parameter $\ell_o:=V_o^{1/3}$, and denote the ratio of $\ell_o$ to $a_o$ by $\vartheta$ which is equal to $(2\pi^2)^{1/3}$. Note that,
If we put $\vartheta$ equal to 0, the equations for the $k$=1 model reduce to those for
the $k$=0 model.

The connections and densitized triads in terms of fiducial quantities are given by
\be
A_a^i=\frac{c}{\ell_o}{}^o\omega_a^i\ \ \ \textrm{and}\ \ \ E_i^a=\frac{p}{\ell_o^2}\sqrt{{}^oq}{}\ ^oe_i^a \, .
\ee
The fundamental Poisson bracket is then
\be
\{c,p\}=\frac{8\pi G\gamma}{3}\, ,
\ee
where $\gamma$ is the Barbero-Immirizi parameter.
At each point $(c,p)$ of the gravitational phase space, the physical 3-metric $q_{ab}$ is given by
\be
q_{ab}=\frac{|p|}{\ell_o^2}{}^oq_{ab}\, .
\ee
Therefore, the physical volume $V$ of the universe, corresponding to this metric is
$V=|p|^{3/2}$.

Since the fiducial frames and co-frames are fixed, and due to the parametrization of connections and triads we have chosen, the only remaining constraint is given by the Hamiltonian constraint which has the form
\be\label{FHC}
\mathcal C_H=\int_\mathcal V N\left[-\frac{E_i^aE_j^b}{16\pi G\gamma^2\sqrt{|q|}}\epsilon^{ij}_{\ k}\left(
F_{ab}^k-(1+\gamma^2)\Omega_{ab}^k\right)+\mathcal H_{\mathrm{matter}}\right]\textrm{d}^3x\, ,
\ee
where $N$ is the lapse function, $\mathcal{H}_{\mathrm{matter}}=\rho V$ and $\Omega_{ab}$ is the curvature of the spin connection $\Gamma_a^i$ which is compatible with the triads. For the $k$=1 model we have,
\be
\Omega^k_{ab}=-\frac{1}{a_o^2}\,\epsilon^k_{\ ij}{}^o\omega_a^i{}^o\omega_b^j
\ee
With this we end the classical preliminaries of this model. Let us now consider the quantization of the system.

\section{Loop Quantization}
\label{sec:2}

To quantize the model, we have to define an appropriate constraint operator. For that we should introduce well defined operators related to each term of it. In the full theory the elementary variables are the holonomies $h_i^{(\mu)}$ along edge $\mu{}^oe_i$ defined by the connection $A^i$, and the flux variables represented by $p$. We need to express the constraint in terms of these functions and their Poisson brackets and after that, replace them by the operator $\hat {h}_i^{(\mu)}$ and $\hat p$ and their commutators divided by $i\hbar$. To find a well defined operator related to the term $\epsilon^{ij}_{\ k}E_i^aE_j^b/\sqrt{|q|}$ we can use Thiemann's strategy. We have the following classical identity from the full theory
\be\label{CTI}
{\epsilon^{ij}}_{k}\,\frac{E_i^aE_j^b}{\sqrt{|q|}}=\frac{\varepsilon}{2\pi G\gamma\mu}\ ^o\epsilon^{abc} \ ^o\omega^i_c\textrm{Tr}\left(h_i^{(\mu)}\{h_i^{(\mu)-1},V\}\tau_k\right)\, ,
\ee
where $\varepsilon$ is 1 if the fiducial and physical frames have the same orientation, otherwise it is -1. Note that here, the parameter $\mu$ is arbitrary, so we have and {\em exact} identity for any (positive) value of $\mu$. Let us now consider the corresponding operator given by
\be\label{QTE}
\widehat{{\epsilon^{ij}}_{k}\,\frac{E_i^aE_j^b}{\sqrt{|q|}}} :=
\frac{\varepsilon}{2i\hbar\pi G\gamma\mu}\ ^o\epsilon^{abc} \ ^o\omega^i_c\textrm{Tr}\left(\hat h_i^{(\mu)}[\hat h_i^{(\mu)-1},\hat V]\tau_k\right)\, ,
\ee
where $\hat V=|\hat p|^{3/2}$ is the volume operator. One should note that, contrary to the classical identity, for different choices of $\mu$, we will have different operators which are {\em not} equivalent.

If we choose the matter field content of the theory to be a massless scalar field, we will also have a term $V^{-1}$ in the constraint. To introduce a well defined operator for it, we use the same strategy as above. Classically, there is the following identity,
\be
V^{-1}=\left(\frac{3}{8\pi G\gamma\mu\ell j(j+1)(2j+1)}\textrm{Tr}(h_i^{(\mu)}\{h_i^{(\mu)-1},V^{2\ell/3}\}\tau_i)\right)^{3/(2-2\ell)}\, ,
\ee
where $\ell$ is a number between 0 and 1 and $j\in \frac{1}{2}\mathds{N}$ is a label for the representation in which the holonomy is evaluated.
Therefore, the corresponding operator will be
\be\label{IV}
\widehat{V^{-1}} := \left(\frac{3}{8i\hbar\pi G\gamma\mu\ell j(j+1)(2j+1)}\textrm{Tr}(\hat h_i^{(\mu)}[\hat h_i^{(\mu)-1},\hat V^{2\ell/3}]\tau_i)\right)^{3/(2-2\ell)}\, .
\ee
We should note that with different choices of $\ell$, $j$ and $\mu$, we  have different operators but they approach each other when $V$ goes to infinity. For simplicity, we shall take $j=1/2$, $\ell=1/2$ or $1/4$.

To define the operator corresponding to the curvature $F_{ab}$, there are two possibilities. The first option, which we shall call the {\it curvature based quantization} is finding $F_{ab}$ by calculating a holonomy along a square loop with physical area equal to the smallest eigenvalue of area operator in the full theory, hence
 \be\label{CuC}
 \hat F^k_{ab}=\frac{\varepsilon}{\bar\mu^2\ell_o^2}(\sin^2\bar\mu(c-\vartheta)-\sin^2\bar\mu\vartheta)\ ^o\epsilon^k_{\ ij}\ ^o\omega_a^i\ ^o\omega_b^j\, .
 \ee

The other possibility, which we call {\it connection based quantization} is defined by specifying the connection by calculating an open holonomy along a curve which starts and ends at some point $x$. It is made of two paths. One is parallel to one of the fiducial frames and the other is antiparallel, and both have the same physical length equal to the square root of smallest eigenvalue of area operator. One then calculates $F_{ab}$ by using this connection. Therefore, for this prescription the curvature becomes
 \be\label{CoC}
 \hat F_{ab}^k=\frac{1}{\ell_o^2}\,\left(\frac{\varepsilon}{\bar\mu^2}\sin^2\bar\mu c-\frac{2\vartheta}{\bar\mu}\sin\bar\mu c \right)\ ^o\epsilon^k_{\ ij}\ ^o\omega_a^i
 \ ^o\omega_b^j\, .
 \ee
In Eqs.~(\ref{CuC}) and (\ref{CoC}), we have made the choice $\bar\mu=\lambda/\sqrt{|p|}$,
 where $\lambda^2$ is equal to smallest eigenvalue of the area operator.

Since the length of each path used for calculating the holonomies needed to find the curvature $F_{ab}$ in the fiducial cell is $\bar\mu\ell_o$, we will use this length to compute operators
in Eqs.~(\ref{QTE}) and (\ref{IV}).
For example, to find $\widehat{V^{-1}}$, if we choose $\ell=1/2$ then
$\widehat{V^{-1}}=\hat f^3$ and $\hat f\psi(V)=f(V)\psi(V)$ where
\be\label{fv}
f(V)=\frac{3}{2V_c}V^{1/3}\,\left[(V+V_c)^{1/3}-|V-V_c|^{1/3}\right]
\ee
where $V_c=2\pi\gamma\lambda\ell_{\mathrm{Pl}}^2$. But if we choose $\ell=1/4$ then $\widehat{V^{-1}}=\hat g^6$, where the eigenvalues of $\hat g$ are given by
\be\label{gv}
g(V)=\frac{V^{1/3}}{V_c}\,\left[\sqrt{V+V_c}-\sqrt{|V-V_c|}\right]
\ee

We take the gravitational part of the kinematical Hilbert space to be equal to the square integrable functions on the Bohr compactification on the real line, $L^2(\R_{\mathrm{Bohr}},\mathrm{d} \mu _{\mathrm{Bohr}})$. In that space, we can choose a basis of eigenstates,
\be
\hat v|v\rangle=v|v\rangle
\ee
which is related to the volume operator $\hat{V}$ as follows: $\hat{V}|v\rangle=\left(\f{8\pi\gamma}{6}\right)^{3/2}\f{|v|}{K}|v\rangle=V_c|v||v\rangle$ with
$K=2\sqrt{2}/(3\sqrt{3\sqrt{3}})$ and $V_c=2\pi\gamma\lambda\ell_{\textrm{Pl}}^2$.
This space is endowed with the hermitian scalar product defined by
\be
\langle v_i|v_j\rangle=\delta_{v_i,v_j}.
\ee

In this basis, the action of the operator $\widehat{e^{i\alpha\frac{\bar\mu c}{2}}}$ is given by
\be
\widehat{e^{i\alpha\frac{\bar\mu c}{2}}}|v\rangle=|v-\alpha\rangle\ \ \ , \ \ \alpha\in\R.
\ee

In the next two sections we shall study both the curvature and connection based quantizations to find modified effective equations which incorporate the effects of inverse triads operators. Since an operator such as $\hat{f}$ which is a function of $\hat p$ is not the inverse of $\widehat {f^{-1}}$, we shall not ignore any negative power of $p$ and do not cancel them with positive ones before quantization. Furthermore, in the rest of paper, we shall choose lapse function $N=1$. To quantize the constraint with lapse function which differs from 1, we can quantize the constraint with lapse equals to one and find a well defined operator for lapse function and order them.
With this, we can then arrive at the quantum operator for both types of quantization.

For the curvature based quantization,
the constraint operator, without operator ordering, can be written as,
\be
\hat{\mathcal C}_H = -\frac{3}{8\pi G\gamma^2\lambda^2}\hat A\,(\hat{|p|}^{3/2}\sin^2\bar\mu(c-\vartheta)-\hat{|p|}^{3/2}\sin^2\bar\mu\vartheta+(1+\gamma^2)\lambda^2\vartheta^2\hat{|p|}^{1/2})+\hat\rho\,\hat{|p|}^{3/2}
\ee
where
\be
\hat A=\frac{1}{2V_c}\bigg(\frac{3}{4\pi\gamma}\bigg)^{3/2}\bigg(\widehat{e^{-i\frac{\lambda\beta}{2}}}\hat V\widehat{e^{i\frac{\lambda\beta}{2}}}-\widehat{e^{i\frac{\lambda\beta}{2}}}\hat V\widehat{e^{-i\frac{\lambda\beta}{2}}}\bigg)\, .
\ee
is a part of the operator in eq.(\ref{QTE}). The eigenvalues of this operator in term of volume are
\be\label{Av}
A(V)=\frac{1}{2V_c}(V+V_c-|V-V_c|)=
\left\{\begin{array}{lr}V/V_c & V< V_c\\1 & V\geq V_c\end{array}\right.
\ee

For the connection based quantization, the constraint operator without factor ordering, is
given by,
\be
\hat{\mathcal C}_H = -\frac{3}{8\pi G\gamma^2\lambda^2}\hat A\,(\hat{|p|}^{3/2}\sin^2\bar\mu c-2\lambda\vartheta\hat{|p|}\sin\bar\mu c+(1+\gamma^2)\lambda^2\vartheta^2\hat{|p|}^{1/2})+\hat\rho\,\hat{|p|}^{3/2}
\ee
Because of the form of the operators contained in the constraint operator, it is simpler to work with variables $\beta$ and $V$ and take $\hat V$ and $\widehat{e^{i\lambda\beta/2}}$ as elementary operators.
If we order operators the same as in previous works on $k$=1 model with curvature based quantization in $\bar\mu$ scheme \cite{apsv}, the Hamiltonian constraint operator has the form
\bea
\hat{\mathcal C}_{H}^{(1)}&=&-\frac{3}{8\pi G\gamma^2\lambda^2}\bigg(\widehat{\sin(\lambda\beta)}\hat V\hat A\,\widehat{\sin(\lambda\beta)}-\lambda\vartheta\bigg[\widehat{\sin(\lambda\beta)}\hat V^{2/3}\hat A+\hat V^{2/3}\hat A\,\widehat{\sin(\lambda\beta)}\bigg]\nonumber\\
&&+\lambda^2\vartheta^2(1+\gamma^2)\hat V^{1/3}\hat A\bigg)+\hat{\mathcal C}_{\mathrm{matt}},
\label{C1}
\eea
The final quantum theory has a structure very similar to that of \cite{apsv}.
An interesting state for consideration is the one which has support on zero volume namely $\Psi(0;\phi)$. Somewhat intuitively, since this state is an eigenstate of volume
with zero eigenvalue, one could identify this state with the physical singularity.
While the issue of singularity resolution is much more subtle (see, for instance \cite{slqc,aa:ps} for discussions), it is still instructive to study the behaviour of such state. For instance, the left side of Eq.(\ref{C1}) annihilates this state but the right hand side has a non trivial action. This implies that the contribution from $\Psi(0;\phi)$
has to vanish, so there is no `zero-volume' component of any admissible physical state.

In order to write the complete constraint we need an expression for the matter part. For the massless scalar field we have
\be
\hat{\mathcal C}_{\mathrm{matt}}^{(1)}=\frac{\hat{p_\phi}^2}{2}\;\hat V\,\widehat{V^{-2}},
\ee
where the action of $\widehat{V^{-2}}$ is defined as follows
\be
\widehat{V^{-2}}|v\rangle=g^{12}(v)|v\rangle\quad ,\quad g(v)=\frac{|v|^{1/3}}{V_c^{1/6}}\bigg|\sqrt{|v+1|}-\sqrt{|v-1|}\bigg|.
\ee
Therefore, the action of the total Hamiltonian constraint operator
$$\hat{\mathcal C}^{(1)}\cdot\Psi(v,\phi)=(\hat{\mathcal C}_{\mathrm{grav}}^{(1)}+\hat{\mathcal C}^{(1)}_{\mathrm{matt}})\Psi(v,\phi)=0,$$ on a state $\Psi(v,\phi)$ is given by
\bea
-\partial_\phi^2\Psi(v,\phi)&=&-\frac{3}{4\pi G\gamma^2\lambda^2}g(v)^{-12}V(v)^{-1}\Bigg[\bigg(V(v+2)A(v+2)+V(v-2)A(v-2)\nonumber\\
&&+\lambda^2\vartheta^2(1+\gamma^2)V^{1/3}(v)\bigg)\Psi(v,\phi)\nonumber\\
&&+\frac{\lambda\vartheta}{2i}\bigg(\sgn(v)V^{2/3}(v)A(v)+\sgn(v-2)V^{2/3}(v-2)A(v-2)\bigg)\Psi(v-2,\phi)\nonumber\\
&&-\frac{\lambda\vartheta}{2i}\bigg(\sgn(v)V^{2/3}(v)A(v)+\sgn(v+2)V^{2/3}(v+2)A(v+2)\bigg)\Psi(v+2,\phi)\nonumber\\
&&-V(v-2)A(v-2)\Psi(v-4,\phi)-V(v+2)A(v+2)\Psi(v+4,\phi)\Bigg]
\eea

As is common with any quantization process, there are several factor orderings possible (for a discussion of different choices in the $k$=0 model, see \cite{madrid}). In our case, one can define a second constraint operator given by,
\bea
\label{CH2}
\hat{\mathcal C} _{H}^{(2)}&=&\hat{\mathcal C}_{\textrm{grav}}^{(2)}+\hat{\mathcal C}_{\textrm{matt}}^{(2)}\nonumber\\
&&-\frac{3}{8\pi G\gamma^2\lambda^2}\bigg[\hat V^{1/3}\sin\lambda\beta\hat V^{1/3}\hat A\sin\lambda\beta\hat V^{1/3}-\lambda\vartheta\hat V^{1/3}(\hat\epsilon\hat A\sin\lambda\beta\hat+\sin\lambda\beta\hat\epsilon\hat A) V^{1/3}\nonumber\\
&&+\lambda^2\vartheta^2(1+\gamma^2)\hat V^{1/3}\hat A\bigg]+\frac{\hat p_{\phi}^2}{2}\hat V^{-2}\hat V
\eea
The action of this Hamiltonian operator on a state is given by
\be
\label{C2}
-g(v)^{12}V(v)\partial_{\phi}^2\Psi(v;\phi)=\hat{\mathcal C}_{\textrm{grav}}^{(2)}\Psi(v;\phi)
\ee
where
\bea
\label{CA2}
\hat{\mathcal C}_{\textrm{grav}}^{(2)}\Psi(v;\phi)&=&-\frac{3}{8\pi G\gamma^2\lambda^2}\bigg[-V^{1/3}(v)V^{1/3}(v+2)V^{1/3}(v+4)A(v+2)\Psi(v+4;\phi)\nonumber\\
&&+\frac{i\lambda\vartheta V^{1/3}(v)V^{1/3}(v+2)}{2}[\textrm{sgn}(v)A(v)+\textrm{sgn}(v+2)A(v+2)]\Psi(v+2;\phi)\nonumber\\
&&+V^{2/3}(v)[V^{1/3}(v-2)A(v-2)+V^{1/3}(v+2)A(v+2)\nonumber\\
&&+\lambda^2\vartheta^2(1+\gamma^2)V^{1/3}(v)A(v)]\Psi(v;\phi)\nonumber\\
&&-\frac{i\lambda\vartheta V^{1/3}(v)V^{1/3}(v-2)}{2}[\textrm{sgn}(v)A(v)+\textrm{sgn}(v-2)A(v-2)]\Psi(v-2;\phi)\nonumber\\
&&-V^{1/3}(v)V^{1/3}(v-2)V^{1/3}(v-4)A(v-2)\Psi(v-4;\phi)\bigg]
\eea
In this case, $\hat{\mathcal C}_{\textrm{grav}}^{(2)}$ annihilates the $\Psi(0;\phi)$ state, so it does not evolve into the other states. On the other hand, the only states which can evolve to this state are $\Psi(0\pm2;\phi)$ and $\Psi(0\pm4;\phi)$ but in that case, their coefficients are zero. This means that $\Psi(0;\phi)$ is an `isolated' state which cannot evolve to other states and any of other states cannot evolve into this one. Therefore, with this choice (and other similar choices) of operator ordering, it is easy to see that the singularity is resolved at quantum level.

One should note that the factor orderings here chosen differ slightly from that in \cite{ck2}, where the factor ordering was selected to be closer to the Bianchi II and IX cases available in the literature \cite{bianchiII,bianchiIX}.
Let us now consider some properties of this  Hamiltonian constraint. In particular, we shall
explore the issue of self-adjointness of the quantum operator.

\section{Properties of Gravitational part of the Hamiltonian Constraint: Self Adjointness}
\label{sec:3}

In this section we shall focus our attention on the self-adjointness of the quantum constraint operators for the connection based quantization with inverse volume corrections. We shall focus our attention on the operator $\hat{\cal C}^{(1)}$ defined above, and shall briefly comment on the other possible operator $\hat{\cal C}^{(2)}$.
In a sense
the results here presented can be seen as an extension of the results of \cite{closed} to the new quantization and with the inclusion of inverse corrections.

As we  mentioned in previous section, the gravitational part of kinematical Hilbert space is Cauchy completion of the vector space of formal finite linear combinations of the elements of the basis $\{|v\rangle:v\in\R\}$ in which the volume operator (and $\hat p$) is diagonalized.

We recall that the gravitational part of Hamiltonian constraint, when the lapse function $N$ is equal to 1, is given by
\bea
\hat{\mathcal C}_{\mathrm{grav}}^{(1)}&=&-\frac{3}{8\pi G\gamma^2\lambda^2}\bigg(\widehat{\sin(\lambda\beta)}\hat V\hat A\,\widehat{\sin(\lambda\beta)}-\lambda\vartheta\bigg[\widehat{\sin(\lambda\beta)}\hat V^{2/3}\hat A+\hat V^{2/3}\hat A\,\widehat{\sin(\lambda\beta)}\bigg]\nonumber\\
&&+\lambda^2\vartheta^2(1+\gamma^2)\hat V^{1/3}\hat A\bigg),
\nonumber
\eea
The domain of this operator is
\be
D=\{\psi\in\mathcal H^{\mathrm{kin}}/\psi=\sum_{i=1}^n a_i|v_i\rangle,\ a_i\in\C,\ v_i\in\R,\ n\in\mathbb N \}.
\ee
We can express the action of the operators contained in the scalar constraint operator in this domain:
\bea
\hat A|v\rangle&=&A(v)|v\rangle,\ \ \ \ A(v)=\frac{1}{2}(|v+1|-|v-1|),\\
\widehat{\sin(\lambda\beta)}|v\rangle&=&\frac{1}{2i}(|v-2\rangle-|v+2\rangle).
\eea
The operator $\hat{\mathcal C}_{\mathrm{grav}}^{(1)}$ can then be written in a simpler form as
\be
\hat{\mathcal C}_{\mathrm{grav}}^{(1)} = -\frac{3}{8\pi G\gamma^2\lambda^2}(C_0+C_{+2}U_{+2}+C_{-2}U_{-2}+C_{+4}U_{+4}+C_{-4}U_{-4}),
\ee
where $U_i$, $i=\pm2,\pm4$, is the shift operator defined as
$$U_i|v\rangle=|v+i\rangle$$
and $C$'s are some functions of the variable $v$ given by
\bea
\label{cfunc}
C_0(v)&=&V(v+2)A(v+2)+V(v-2)A(v-2)+\lambda^2\vartheta^2(1+\gamma^2)V^{1/3}(v)A(v),\\
C_{-2}(v)&=&-\frac{\lambda\vartheta}{2i}[V^{2/3}(v)A(v)+V^{2/3}(v-2)A(v-2)],\\
C_{+2}(v)&=&\frac{\lambda\vartheta}{2i}[V^{2/3}(v)A(v)+V^{2/3}(v+2)A(v+2)],\\
C_{-4}(v)&=&-V(v-2)A(v-2),\\
C_{+4}(v)&=&-V(v+2)A(v+2).
\eea
In this form, it is clear that $\hat{\mathcal C}_{\mathrm{\mathrm{grav}}}$ preserves the subspaces
\be
\mathcal H_\epsilon=\textrm{Span}\ \bigg(|2n+\epsilon\rangle:\ \ n\in\mathbb Z\ \ ,\ \ \epsilon\in[0\ ,\ 2)\bigg).
\ee
The kinematical Hilbert space can then be decomposed into these preserved subspaces
\be
\mathcal H_{\mathrm{kin}}=\overline{\bigoplus_\epsilon \mathcal H_\epsilon}.
\ee
In what follows, we shall prove the following properties for the operator $\hat{\mathcal C}_{\mathrm{grav}}^{(1)}$:
\begin{itemize}
\item[1.] $\hat{\mathcal C}_{\mathrm{grav}}^{(1)}$ is essentially self adjoint.
\item[2.] $\hat{\mathcal C}_{\mathrm{grav}}^{(1)}$ is sharply negative.
\item[3.] For each of the subspaces $\mathcal H_\epsilon$,  the restricted operator $\hat{\mathcal C}_{\mathrm{grav}}^{(1)}:
\mathcal H_\epsilon\rightarrow \mathcal H_\epsilon$ considered as an essentially self adjoint operator in the Hilbert space
$\overline{\mathcal H_\epsilon}$ has a discrete spectrum.
\item[4.] The above restricted operator satisfies:
 $$ \textrm{dim}\mathcal
H_{\mathcal C_{\mathrm{grav}}^{(1)}>-E}\le\textrm{dim}\mathcal H_{A' \ge-E}$$
for arbitrary $E>0$ where
\be
A'\ :=\ - \frac{3\vartheta^2}{8\pi G}\hat V^{1/3}\hat A\ :\ \ \mathcal
H_\epsilon\rightarrow\
\mathcal H_\epsilon\, .
\ee
\end{itemize}
Let us now prove all these statements.

\bigskip

\noindent
{\bf Property 1.} To prove this property we use the following theorem (theorem VIII.3 of \cite{RS}):
\medskip

{\bf Theorem} Let $T$ be a densely defined symmetric operator on a Hilbert space. Then the followings are equivalent:
\begin{itemize}
\item[a.] $T$ is essentially self-adjoint.
\item[b.] Ker$(T^\ast+z)$ and Ker$(T^\ast+\bar z)$ are equal to $\{0\}$.
\item[c.] Ran$(T+z)$ and Ran$(T+\bar z)$ are dense.
\end{itemize}
Therefore, it is sufficient to show that Ker$(\hat{\mathcal C}_{\mathrm{grav}}^\ast\pm i)$ is equal to the set $\{0\}$. \\
Let $\phi\in D(\hat{\mathcal C}_{\mathrm{grav}}^\ast)$ be such that
\be
\label{ACE}
\hat{\mathcal C}_{\mathrm{grav}}^\ast\phi=\mp i\phi.
\ee
\\
The matrix representation of the operator $\hat{\mathcal C}_{\mathrm{\mathrm{grav}}}^{(1)}$ can be defined in the basis $\{|v\rangle | v\in\mathbb R\}$ as
$$A_{vw}=\langle v|\hat{\mathcal C}_{\mathrm{grav}}^{(1)}|w\rangle,$$
and since $\hat{\mathcal C}_{\mathrm{grav}}^{(1)}$ is symmetric, $A_{vw}=A^\ast_{wv}$. Note that each row and column of $A$ has, at most, five nonzero elements.\\
{\bf Lemma}
Let $T$ be an operator on the Hilbert space $\mathcal H^{\mathrm{kin}}$ with domain $\mathcal D$ equal to $D(\hat{\mathcal C}_{\mathrm{grav}}^{(1)})$ and $A=(a_{v,w})_{v,w\in\mathbb R}$ be its matrix representation. Then, for $\phi$ and $\psi$ in $\mathcal H^{\mathrm{\mathrm{kin}}}$, the following statements are equivalent \cite{JP}
\begin{itemize}
\item[i.]
$\langle v, \phi\rangle=\sum_{w}a_{v,w}^\ast \langle w,\psi\rangle$ is absolutely convergent for each $v\in\mathbb R$ and $\phi\in\mathcal H^{\mathrm{\mathrm{kin}}}$.
\item[ii.]
$\psi\in D(T^\ast)$ and $T^\ast\psi=\phi$.
\end{itemize}
By applying the above lemma, eq.(\ref{ACE}) can be written as
\be
\sum_{v}\sum_w a_v A_{v,w} |v\rangle=\mp i\sum_v a_v|v\rangle.
\ee
 After multiplying $\phi$ from the left on both sides, the above equation gets the form
\be
\sum_{v}\sum_w a_w a^\ast_v A_{v,w}=\mp i\sum_v |a_v|^2.
\ee
Since the left hand side is a real number and the right hand side is an imaginary number, $|\phi|$ and therefore $\phi$ should be 0 and the proof is complete.


One important feature is that it is indeed possible to generalize the above proof so that it applies to those symmetric operators which are defined on a Hilbert space $\mathcal H$ isomorphic to $L^2(S,\mu)$, where $S$ is some measure space with counting measure $\mu$ and with domain $\mathcal D$  a subset of $\mathcal H$ which contains all finite linear combinations of the elements of the orthogonal basis of $\mathcal H$. Furthermore, the index set can be countable or uncountable, and their matrix representation has the same properties as the matrix of $\hat{\mathcal C}_{\mathrm{grav}}^{(1)}$, which means that in each row and column they have a finite number of elements.
\\
The gravitational part of the Hamiltonian operators for the flat FRW model, closed FRW with curvature based quantization, Bianchi I, II and IX which come from LQC are some examples of theories where the corresponding geometric constraint operators are similar to $\hat{\mathcal C}_{\mathrm{grav}}$, so  that one can apply the general form of the above proof to show that they are essentially self adjoint in their domain.
\bigskip\\
\noindent{\bf Property 2.}
The operator $\hat{\mathcal C}_{\mathrm{grav}}^{(1)}$ can be written as
\be
\hat{\mathcal C}_{\mathrm{grav}}^{(1)}=-\frac{3}{8\pi G\gamma^2\lambda^2}\bigg(\widehat{\sin(\lambda\beta)}\hat V^{1/3}-\lambda\vartheta\bigg)\hat A\hat V^{1/3}\bigg(\widehat{\sin(\lambda\beta)}\hat V^{1/3}-\lambda\vartheta\bigg)^\dag-\frac{3\vartheta^2}{8\pi G}\hat A\hat V^{1/3}.
\ee
From the following inequalities
\be
\bigg(\widehat{\sin(\lambda\beta)}\hat V^{1/3}-\lambda\vartheta\bigg)\hat A\hat V^{1/3}\bigg(\widehat{\sin(\lambda\beta)}\hat V^{1/3}-\lambda\vartheta\bigg)^\dag\geq 0,
\ee
\be
\frac{3\vartheta^2}{8\pi G}\hat A\hat V^{1/3}\geq 0,
\ee
it is obvious that $\hat{\mathcal C}_{\mathrm{grav}}^{(1)}$ is negative definite and satisfies,
\be
\hat{\mathcal C}_{\mathrm{grav}}^{(1)}\leq-\frac{3\vartheta^2}{8\pi G}\hat A\hat V^{1/3}\leq 0.
\ee
As a stronger result, we now show that the equation $\hat{\mathcal C}_{\mathrm{grav}}^{(1)}\phi=0$ has no nontrivial solution for $\phi$ and hence $\hat{\mathcal C}_{\mathrm{grav}}^{(1)}<0$.\\
Assume some $\phi=\sum_{j=1}^n a_j|v_j\rangle$ as an element of $D$ be a solution of equation $\hat{\mathcal C}_{\mathrm{grav}}^{(1)}\phi=0$. Thus, we have
\bea
0&=&\sum_{j=1}^na_j\bigg(C_0(v_j)|v_j\rangle+C_{-2}(v_j)|v_j-2\rangle+C_{+2}(v_j)|v_j+2\rangle\nonumber\\
&&+C_{-4}(v_j)|v_j-4\rangle+C_{+4}(v_j)|v_j+4\rangle\bigg).
\eea
If there exists a non trivial $\phi$ that satisfies the above equation, the terms with the same $|v\rangle$ should cancel each other. Without loss of generality, one can write $\phi=\sum_{j=1}^n a_j|v_j\rangle$ in such a way that $v_1\leq v_2\leq...\leq v_n$. Hence, it is easy to see that there are two terms $a_nC_{+4}(v_n)|v_n+4\rangle$ and $a_1C_{-4}(v_1)|v_1-4\rangle$ which cannot cancel with other terms and the only way to cancel them is to put $a_1C_{-4}(v_1)=0$ and $a_nC_{+4}(v_n)=0$. According to the definition of $C_{-4}$ and $C_{+4}$ and our assumption $v_1\leq v_n$, $C_{-4}$ and $C_{+4}$ cannot be zero simultaneously and therefore at least one of $a_1$ or $a_n$ should be zero. This means there is no term with $|v_1\rangle$ or $|v_n\rangle$ in $\phi$. We can repeat the same argument and show step by step that all $a_j$'s in $\phi$ are zero and therefore $\phi=0$.\\
\bigskip\\
{\bf Properties 3 and 4}
Here we shall use the same procedure as in Ref. \cite{closed}.
\medskip\\
{\bf Definition:} Let $T$ be an essentially self-adjoint operator defined in some
domain $\mathcal D$ in the Hilbert space ${\mathcal H}$. We say its
spectrum is discrete whenever the following conditions are
satisfied:
\begin{itemize}
\item there exists a basis of ${\mathcal H}$ consisting of the eigenvectors of $T$,
\item for each eigenvalue the corresponding eigenvectors  span a finite dimensional subspace,
\item for every finite interval $I$ of $\R$, the set of the eigenvalues
of $T$ contained in $I$ is finite.
\end{itemize}
Also let $T$ be a self-adjoint operator in a Hilbert space $\mathcal H$, $\alpha$ be a real number and $\iota$ be an inequality relation, we define ${\mathcal P}_{T\iota\alpha}:{\mathcal H}\rightarrow{\mathcal H}$ as the spectral projector of $T$ onto the interval
$\{t\in\R\ :\ t\iota\alpha\}$. We mean by $\mathcal H_{T\iota\alpha}$ to be the image of $\mathcal P_{T\iota\alpha}$
\be
{\mathcal H}_{T\iota\alpha}:=\mathcal P_{T\iota\alpha}(\mathcal H).
\ee
To prove the properties 3 and 4, we use the following lemma \cite{closed}:
\medskip\\
{\bf Lemma:} Let ($B$,$\mathcal D(B)$) and ($B'$,$\mathcal
D(B')$) be operators in a Hilbert space $\mathcal H$ with their
domains and $\mathcal D\subset \mathcal D(B)\cap \mathcal D(B')$ be a dense
subspace of $\mathcal H$.
Suppose the following conditions are satisfied:
\begin{itemize}
\item On the domain $\mathcal D$ the following inequality holds
$$ 0\le B\le B',$$
\item The operator $B'$ is essentially self-adjoint in $\mathcal D$,
\item $B$, as an operator defined in $\mathcal D(B)$, is self adjoint,
positive and has discrete spectrum.
\end{itemize}
Then $B$ is also positive and has discrete spectrum. Moreover, the
following inequality holds for arbitrary $\lambda\ge 0$
\[
\textrm{dim}\mathcal H_{B'<\lambda}\ \le\ \textrm{dim}\mathcal
H_{B\le\lambda}.
\]
By fixing $\epsilon\in\R$ and defining  $\mathcal H=\overline{\mathcal H_\epsilon}$ and $\mathcal D=\mathcal H_\epsilon$ as the corresponding Hilbert space and the domain of the following operators
\begin{align}
B=\hat A\,\hat V^{1/3}\ &:\ \mathcal H_\epsilon\rightarrow\mathcal H_\epsilon,\\
B'=-\frac{8\pi G}{3\vartheta^2}\,\hat{\mathcal C}_{\mathrm{grav}} \ &:\ \mathcal H_\epsilon\rightarrow\mathcal H_\epsilon,
\end{align}
the above lemma implies the properties 3 and 4.

Let us now briefly summarize the situation for the operator $\hat{\mathcal C}_{\mathrm{grav}}^{(2)}$. It is straightforward to show that this operator satisfies
the properties 1, 3 and 4 above. It does not satisfy property 2, since the operator is only
non-positive.

One can conclude that these results  establish the  self-adjointness nature of the quantum operators introduced in the previous Section. An interesting spin-off of the results here proved is that they imply also the self-adjointness of the connection based quantization without inverse corrections, as defined in \cite{ck2}, as well as for the curvature based operator with inverse corrections. As we have mentioned above, some of the methods of proof can be readily applied to other operators of interest in anisotropic loop quantum cosmology.

Let us now analyse the effective theories defined by
both quantum theories. In the next Section we shall first consider the `curvature based' quantization followed by the study of the connection based.


\section{Effective Theory}
\label{sec:4}

As is now usual practice in LQC, the study of the effective description for the model in
question
can provide valuable information regarding the dynamics of certain semiclassical states
(for a detailed description of the motivation and derivation of such description see
\cite{aa:ps} and \cite{ia:ac}). This section has two parts. In the first one we consider
the theory obtained by means of the curvature based quantization. In the second part, e consider the theory obtained by means of the connection based quantization.

\subsection{Curvature Based Quantization}

Let us first consider the theory for which the curvature is obtained via closed holonomies.
In this case, it is straightforward to see that the effective Hamiltonian is
\be
\begin{split}
\mathcal{H}_{\mathrm{eff}}=&-\frac{3}{8\pi G\gamma^2\lambda^5}\left[|p|^{3/2}\sin^2\bar\mu(c-\vartheta)- |p|^{3/2}\sin^2\bar\mu\vartheta+(1+\gamma^2)\lambda^2\vartheta^2\sqrt{|p|}\right]\\
&+\rho |p|^{3/2}
\end{split}
\ee
Since we want to find some effective equations which give us a possibility to study the behavior of the system for small volumes, we have to add the effects of those operators which come from the inverse power of triads. As expected, their eigenvalues have a different behavior than their classical quantities. For this reason we need to use some functions to represent these operators and one natural choice for such functions is to consider the operator's eigenvalues. Therefore, the modified effective equation can be written as
\be
\begin{split}
\mathcal{H}_{\mathrm{eff}}=&-\frac{3}{8\pi G\gamma^2\lambda^5}A(V)\left[|p|^{3/2}\sin^2\bar\mu(c-\vartheta)- |p|^{3/2}\sin^2\bar\mu\vartheta+(1+\gamma^2)\lambda^2\vartheta^2\sqrt{|p|}\right]\\
&+\rho |p|^{3/2}
\end{split}
\ee
Now, we can rewrite the above equation in terms of the new variables $\beta=c/\sqrt{|p|}$ and $V=|p|^{3/2}$, with Poisson bracket $\{\beta,V\}=4\pi G\gamma$, as
\be
\begin{split}
\mathcal{H}_{\mathrm{eff}}=&-\frac{3}{8\pi G\gamma^2\lambda^5}A(V)\left[V\sin^2(\lambda\beta-D)-V\sin^2D+(1+\gamma^2)\lambda^2\vartheta^2V^{1/3}\right]\\
&+\rho V\, ,
\end{split}
\ee
where $D=\lambda\vartheta V^{-1/3}$. The equations of motion are then
\be
\dot V=-\{\beta,V\}\frac{\partial\mathcal{H}_{\mathrm{eff}}}{\partial\beta}=\frac{3}{\gamma\lambda}A(V)V\sin(\lambda\beta-D)\cos(\lambda\beta-D)
\ee
and,
\be
\begin{split}
\dot\beta=&\{\beta,V\}\frac{\partial\mathcal{H}_{\mathrm{eff}}}{\partial V}=-\frac{3}{2\gamma\lambda^2}\Bigg[\bigg(VA_{,V}+A(V)\bigg)\bigg(\sin^2(\lambda\beta-D)-\sin^2D\bigg)\\
&+(1+\gamma^2)\lambda^2\vartheta^2A_{,V}V^{1/3}+\frac{\lambda\vartheta}{3}A(V)V^{-1/3}\bigg(\sin2(\lambda\beta-D)+\sin2D\bigg)\\
&+\frac{1+\gamma^2}{3}\lambda^2\vartheta^2A(V)V^{-2/3}\Bigg]+4\pi G\gamma\bigg(V\frac{\partial\rho}{\partial V}+\rho\bigg)
\end{split}
\ee
where the derivative $A_{,V}$ of $A(V)$ is given by
\be
A_{,V}=\frac{1}{2V_c}\left(1-\frac{|V-V_c|}{V-V_c}\right)
=\left\{\begin{array}{lr} 1/V_c & V<V_c\\ 0 & V>V_c\end{array}\right.
\ee
From these equations one can then find the expansion $\theta$ as
\be
\label{exk11}
\theta = \frac{\dot{V}}{V}=\frac{3}{\gamma\lambda}A(V)\sin(\lambda\beta-D)\cos(\lambda\beta-D)
\, .
\ee
Now, it is immediate to see that, since $|\sin2(\lambda\beta-D)|\leq 1$, the expansion is absolutely  bounded. The expansion has some zeros which correspond to turnaround points, which  can be either bounces or recollapses. By using the time derivative of the expansion we can identify the nature of turnaround points \cite{ck2}. The expression for $\dot\theta$ is
\be
\label{dexk11}
\dot\theta=A(V)\cos 2(\lambda\beta-D)\left(\frac{3}{\gamma}\dot\beta+\frac{\theta D}{\gamma\lambda}\right)+\frac{3}{2\lambda\gamma}\dot{V} A_{,V}\sin2(\lambda\beta-D)\, .
\ee
From this expression we can see that, at a turnaround point where $\dot{V}=0$, the second term vanishes and, for $V>V_c$, we are left with an equation very similar to the equation one has when inverse volume effects are neglected \cite{ck2}. In any case, since $A(V)>0$, the nature of the turnaround point is not affected by the inverse corrections.

It is useful to consider the  matter density, and study the modifications it suffers from the inverse corrections. The density is given by
\be\label{RK11}
\rho=\frac{3}{8\pi G\gamma^2\lambda^5}A(V)\left[\sin^2(\lambda\beta-D)-\sin^2D+(1+\gamma^2)\lambda^2\vartheta^2V^{-2/3}\right]\, .
\ee
From this equation one can see that, due to the presence of the factor $A(V)$, if the volume goes to zero, then the matter density goes to zero as well. Furthermore, using  Eqs.~(\ref{Av}) and (\ref{RK11}), it is easy to show that,
\be
\rho\leq\frac{3}{8\pi G\gamma^2\lambda^5}\left[1+(1+\gamma^2)\lambda^2\vartheta^2V_c^{-2/3}\right]\, .
\ee
Therefore, the matter density has a global upper limit, a result which is qualitatively different from the previous case without inverse triad correction \cite{ck2}.

\subsection{Connection Based Quantization}

Let us now consider the effective theory for the other quantization, namely the connection based quantization \cite{ck2}, when inverse volume corrections are considered.
The modified effective Hamiltonian with variables $\beta$ and $V$ is
\be
\begin{split}
\mathcal{H}_{\mathrm{eff}}=&-\frac{3}{8\pi G\gamma^2\lambda^2}A(V)\left[V\sin^2\lambda\beta
-2\lambda\vartheta V^{2/3}\sin\lambda\beta+(1+\gamma^2)\lambda^2\vartheta^2V^{1/3}\right]\\
&+\rho V
\end{split}
\ee
The equations of motion are then,
\be
\dot V=\frac{3}{\gamma\lambda}A(V)\cos\lambda\beta(V\sin\lambda\beta-\lambda\vartheta V^{2/3})\, ,
\ee
and
\be
\begin{split}
\dot\beta=&-\frac{3}{2\gamma\lambda^2}\Bigg[\bigg(VA_{,V}+A(V)\bigg)\sin^2\lambda\beta-2\lambda\vartheta\sin\lambda\beta\bigg(\frac{2}{3}A(V)V^{-1/3}+A_{,V}V^{2/3}\bigg)\\
&+\lambda^2\vartheta^2(1+\gamma^2)\bigg(\frac{A(V)V^{-2/3}}{3}+A_{,V}V^{1/3}\bigg)\Bigg]+4\pi G\gamma\left(V\frac{\partial\rho}{\partial V}+\rho\right)
\end{split}
\ee
The expansion is given by,
\be
\label{exk12}
\theta=\frac{3}{\gamma\lambda}A(V)\cos\lambda\beta(\sin\lambda\beta-\lambda\vartheta V^{-1/3})\, ,
\ee
and its time derivative
\be
\label{dexk12}
\dot\theta=\frac{3}{\gamma}\dot\beta A(V)(\cos2\lambda\beta+D\sin\lambda\beta)+\frac{3}{\gamma\lambda}D\cos\lambda\beta\left(\frac{A(V)\theta}{3}-\dot VA_{,V}\right)\, .
\ee
We can see that the expansion $\theta$ is absolutely bounded because in Eq.~(\ref{exk12}):\\
1. The maximum and minimum values for $\cos\lambda\beta$ are 1 and -1,\\
2. The absolute value of $A(V)(\sin\lambda\beta-\vartheta\lambda V^{-1/3})$ is always less than $1+3^{1/6}\vartheta$,
Therefore, the absolute value of $\theta$ is less than $3/(\gamma\lambda)+3^{7/6}\vartheta/(\gamma\lambda)$.
\par
It should be noted that in the quantization with no inverse corrections \cite{ck2}, the expansion can not be absolutely bounded on the whole (effective) phase space \cite{gupt:singh}. Therefore, a direct consequence of the inclusion of such terms is to render the expansion absolutely bounded, as in the case of the $k$=0 FRW model.

Let us now analyze the behaviour of matter density. First,  we have following expression for density
\be
\rho=\frac{3}{8\pi G\gamma^2\lambda^2}A(V)\left[V\sin^2\lambda\beta
-2\lambda\vartheta V^{2/3}\sin\lambda\beta+(1+\gamma^2)\lambda^2\vartheta^2V^{1/3}\right]\\
\ee
It is now straightforward to see that this expression has an upper limit given by,
\be
\rho\leq\frac{3}{8\pi G\gamma^2\lambda^2}\left[1+2\lambda\vartheta V_c^{-1/3}+(1+\gamma^2)\lambda^2\vartheta^2V_c^{-2/3}\right]\\
\ee
In previous studies with no inverse triad effects it was seen that density for this model is unbounded \cite{ck2}. Just as in the case of the expansion, inclusion of inverse effect implies absolute boundedness of the energy density in the whole phase space.

Let us end this section with some remarks:\\
i) Just as in the case of the curvature based quantization discussed in the previous part, the density goes to zero as volume goes to zero. This feature is qualitatively different from the case without inverse correction, and has some potentially important ramifications.\\
ii) Another common result for both quantization is that, according to Eqs.~(\ref{dexk11}) and (\ref{dexk12}), it can be seen that the nature of the turnaround points only depends on the sign of $\dot\beta$, as discussed in \cite{ck2}. In particular, this implies that in the
evolution there will be more than one type of bounce, a feature not observed in the curvature
based quantization \cite{ck2}. \\
iii) The fact that both the expansion $\theta$ and energy density $\rho$ are absolutely bounded, as is the case in the flat FRLW model, suggests that it might be possible to prove generic non-singularity results along the lines of \cite{cosmos}. Some results, for the curvature based quantization, were already shown in \cite{ps:fv}.\\
iv) One should note that these results regarding the effective theory are, strictly speaking, only an approximation in the following sense. On the one had, in the process of deriving them, we employed the inverse triad corrections which are only important effects for very small volumes. On the other hand, in this regime of small volumes we do not expect to have a good description of the dynamics of semiclassical states (if they exist) in terms of effective equations. It would certainly be worth exploring numerically this regime for exact semiclassical states, and test the validity of the assumptions involved.

\section{Discussion and Conclusions}
\label{sec:5}

Inverse volume corrections represent a unique feature of loop quantization when regularizing the Hamiltonian constraint operator. There has been some controversy regarding the necessity or `utility' of including such effects. Loop quantum cosmology represents a very useful framework to explore these kind of issues since we have full control on the effect of such corrections. This has been the main theme of this article. We know, on the one hand, that such corrections are meaningless for the open flat FLRW model (see for instance \cite{aps2,slqc,aa:ps} for discussions), since in that case the invariance of the classical theory under constant rescalings is incompatible with the appearance of a new length scale, as introduced by such correction terms. In the case of closed universes, on the other hand, one should expect that those terms not only are allowed, but might play an important role in the theory. It has been understood for a long time now that those effects are not relevant (despite claims to the contrary in the literature) for singularity resolution; the so called {\em holonomy corrections} play a much important role in singularity resolution.

What we have seen here is that, for the closed $k$=1 model, the inverse corrections {\em do} play a role when considering the effective dynamics for both curvature and connection based quantizations. In particular, relevant geometrical and matter scalars become absolutely bounded when such effects are taken into account. This is consistent with the observations of \cite{gupt:singh} where it was argued that, for spatially curved models, inverse corrections might be useful to cure this undesired feature. Indeed, for a complete quantization of the Bianchi IX models with inverse corrections, several of the relevant geometrical quantities become better behaved \cite{CKM,ck5}.

One might wonder, for instance, what is the relevance of such results for the model considered here, given that, in any case, all effective trajectories do have a bounce and therefore the singularity is resolved. The answer to this question comes when one is interested in considering properties of the system that involve all possible trajectories on phase space. For instance, if one is interested in computing the probability of inflation for such models, one has to consider all possible trajectories and `weight them' accordingly (along the lines of \cite{inflation0,inflation1}).
If we ignore the inverse volume corrections, the energy density is unbounded on the phase space so the volume under which one has to integrate to compute the probability of inflation becomes infinite and there seems no natural way to regularize it \cite{sloan}. However, when the energy density becomes absolutely bounded, then we have a natural cut-off and one can then hope to compute finite probabilities.
Another possible consequence of the boundedness of the scalars pertains to generic singularity resolution, for generic matter content. It would be interesting to explore those issues along the lines of \cite{ps:fv}. We shall leave these investigations for future publications.

\section*{Acknowledgements}

\noindent We would like to thank E. Montoya, P. Singh and E. Wagner for
discussions and comments. This work was in part supported by CONACyT 0177840 grant, by NSF
PHY 1205388 and by the Eberly Research Funds of Penn State.

\end{document}